\documentclass[aps,prb,twocolumn,showpacs]{revtex4}

\begin{document}

\author{O. A. Tretiakov} 
\author{O. Tchernyshyov} 
\affiliation{Department of Physics and Astronomy, The Johns
Hopkins University, Baltimore, Maryland 21218}

\date{January 23, 2007}

\title{Vortices in thin ferromagnetic films and the skyrmion number}

\begin{abstract}
We point out that a peculiar annihilation of a vortex-antovortex pair observed 
numerically by Hertel and Schneider [Phys. Rev. Lett. 97, 177202 (2006)] 
represents the formation and a subsequent decay of a skyrmion.
\end{abstract}


\maketitle

Creation, annihilation, and fusion of topological solitons is
constrained by conservation of related topological charges.  For
example, in a planar (XY) ferromagnet the destruction of a vortex
always proceeds through its annihilation with an antivortex, a process
that conserves the O(2) winding number.  In this note we highlight the
importance of another topological charge for vortex defects associated
with the three-dimensional nature of the spin.  Even in magnets with
an easy-plane anisotropy the magnetization can and does point out of
the plane at the core of a vortex.\cite{Shinjo00} Even though the core
region is exceedingly small, the direction of its out-of-plane
magnetization $p = \mathrm{sgn}M_z(0)$, henceforth referred to as
polarization, is an important parameter.  For instance, the gyrotropic
force acting on a moving vortex depends on the polarization $p$ but
not on the size of the core,\cite{Thiele73} indicating a topological
nature of the effect.

Recently Hertel and Schneider\cite{HertelSchneider06} performed
numerical simulations of the vortex-antivortex pair annihilation in a
thin magnetic film.  They noted drastically different outcomes for
pairs with parallel and antiparallel core magnetizations.  In the
former case the two defects dissipated quietly, while in the latter
the annihilation was accompanied by a violent burst of spin waves.
Below we show that the difference is due to the conservation of
another topological charge, the skyrmion number.  Hertel and Schneider
observed the formation and decay of a skyrmion.

In a thin film with no intrinsic anisotropy the magnetization is
forced to stay mainly in the plane of the film by dipolar
interactions.  Therefore topological defects in the bulk of the film
are characterized by an O(2) winding number, $n=+1$ for vortices and
$n=-1$ for antivortices.  At the core of these defects, the
magnetization points out of the plane.\cite{Shinjo00}  As a result,
there is a second topological invariant characterizing them, the
skyrmion number
\[
q = \int \frac{d^2r}{8\pi} \, \epsilon_{ij} \, \epsilon_{\alpha\beta\gamma}
\, n_\alpha \, \partial_i n_\beta \, \partial_j n_\gamma 
=  - \int \frac{d^2 r}{4\pi} \, 
\frac{\partial(\cos{\theta},\phi)}{\partial(x,y)},
\]
where $\hat\mathbf{n}(\mathbf r)$ is the unit vector parallel to the
local magnetization $\mathbf{M(r)}$.  Skyrmions were introduced in the
context of the two-dimensional Heisenberg model by Belavin and
Polyakov.\cite{Polyakov} However, similar topological defects were
discussed earlier by Feldtkeller,\cite{Feldtkeller65}
Kleman,\cite{Kleman73} and Thiele.\cite{Thiele73} A vortex with a
winding number $n$ and core polarization $p$ has a half-integer
skyrmion charge\cite{Senthil04} $q=np/2$.  A vortex-antivortex pair
with \textit{parallel} polarizations $p$ have opposite skyrmion
numbers adding to zero and thus belongs to the same topological sector
as uniform ground states.  From the topological perspective, such a
texture can be deformed continuously into a ground state and
apparently this is exactly what happens: the energy decreases
continuously until it reaches the ground-state value.

In contrast, a vortex and an antivortex with \textit{antiparallel}
core polarizations have equal skyrmion numbers adding to a total of
$+1$ or $-1$.  This texture belongs to a nontrivial topological sector
and thus cannot be deformed continuously into a ground state (with
zero skyrmion number).  Eventually the skyrmion decays into spin
waves.  A change in the topological sector requires the injection of a
magnetic monopole,\cite{Senthil04} also known as the Bloch
point.\cite{Thiaville04} This process is strictly forbidden in
continuum theories with a fixed length of magnetization but is allowed
in lattice models.\cite{Pokrovsky98} When the radius of a skyrmion
shrinks to the lattice scale, the skyrmion can decay into spin waves.
While details of this process depend on high-energy physics (lattice
scale or the cost of locally suppressing magnetization length), the
energy of spin waves released in the skyrmion decay can be readily
estimated.

In the continuum approximation with exchange interactions only the
energy is $E = A t \int d^2 r \, |\nabla \hat\mathbf{n}|^2$, where $A$
is the exchange constant and $t$ is the film thickness.  The local
minima of energy in the topological sectors with $q=\pm 1$ are
skyrmion textures with energy\cite{Polyakov} $8\pi A t$.  Since the
dipolar energy $\int d^3r \, \mu_0 |\mathbf{H(r)}|^2/2$ is positive
definite, the exchange part provides a lower bound on the energy of a
vortex-antivortex pair with opposite polarizations.  Furthermore, the
dipolar energy can be neglected when the skyrmion shrinks to the scale
of the exchange length $\lambda = \sqrt{2A/\mu_0M^2}$ (a few
nanometers in permalloy).

For the values of exchange constant $A =1.3\times 10^{-11} \rm{J/m}$
and film thickness $t = 10$ nm used in
Ref.~\onlinecite{HertelSchneider06} we find $E_\mathrm{skyrm} \approx
3.3\times 10^{-18}$ J.  This matches well the energy of the
vortex-antivortex pair just before the explosion $E_\mathrm{pair}
\approx 3.1 \times 10^{-18}$ J (measured from the ground state), see
Fig.~4 in Ref.~\onlinecite{HertelSchneider06}.

\textit{Note added in proof:} We have recently become aware of two
more papers to which the subject of skyrmion-mediated annihilation is
relevant.\cite{LeeChoi05, VanWaeyenberge06}

The authors acknowledge support from the NSF Grant No. DMR-0520491 and
from the Theoretical Interdisciplinary Physics and Astrophysics Center
at JHU.


\begin{thebibliography}{99}

\bibitem{Shinjo00} T. Shinjo, T. Okuno, R. Hassdorf, K. Shigeto, and
T. Ono, Science {\bf 289}, 930 (2000).

\bibitem{Thiele73} A. A. Thiele, \prl \textbf{30}, 230 (1973);
 J. Appl. Phys. \textbf{45}, 377 (1974).

\bibitem{HertelSchneider06} R. Hertel and C. M. Schneider,
Phys. Rev. Lett. \textbf{97}, 177202 (2006).

\bibitem{Polyakov} A. A. Belavin and A. M. Polyakov, Pis'ma ZhETF \textbf{22}, 
503 (1975) [JETP Lett. \textbf{22}, 245 (1975)].

\bibitem{Feldtkeller65} E. Feldtkeller, Z. Angew. Phys.
\textbf{19}, 530 (1965).  

\bibitem{Kleman73} M. Kleman, Phil. Mag. \textbf{27}, 1057 (1973). 

\bibitem{Senthil04} T. Senthil, A. Vishwanath, L. Balents, S. Sachdev,
  and M. P. A. Fisher, Science \textbf{303}, 1490 (2004).

\bibitem{Thiaville04} A. Thiaville, J. M. Garc\'{\i}a, R. Dittrich,
J. Miltat, and T. Schrefl,  \prb {\bf 67}, 094410 (2003).

\bibitem{Pokrovsky98} Ar. Abanov and V. L. Pokrovsky, Phys. Rev. B
\textbf{58}, R8889 (1998).

 \bibitem{LeeChoi05} K.-S. Lee, S. Choi, and S.-K. Kim,
 Appl. Phys. Lett. \textbf{87}, 192502 (2005).

\bibitem{VanWaeyenberge06} B. Van Waeyenberge, A. Puzic, H. Stoll,
K. W. Chou, T. Tyliszczak, R. Hertel, M. F\"ahnle, H. Br\"uckl, K. Rott,
G. Reiss, I. Neudecker, D. Weiss, C. H. Back, and G. Sch\"utz, Nature
(London) \textbf{444}, 461 (2006).

\end{thebibliography}
\end{document}